\documentclass[11pt,twoside]{article}

\usepackage{asp2006}
\usepackage{epsf}

\def\ltsima{$\; \buildrel < \over \sim \;$}
\def\simlt{\lower.5ex\hbox{\ltsima}} % < over ~
\def\gtsima{$\; \buildrel > \over \sim \;$}
\def\simgt{\lower.5ex\hbox{\gtsima}} % > over ~

\begin{document}
\title{Electron Energy Distributions at Relativistic Shock Sites:
Observational Constraints from the Cygnus A Hotspots}  

\author{C.~C. Cheung\altaffilmark{1}, \L. Stawarz}   
\affil{Kavli Institute for Particle Astrophysics and Cosmology, \\Stanford 
University, Stanford, CA 94305, USA}    

\altaffiltext{1}{Jansky Postdoctoral Fellow of the National Radio
Astronomy Observatory}

\author{D.~E Harris}
\affil{Harvard-Smithsonian Center for Astrophysics, \\60 Garden St., 
Cambridge, MA 02138, USA}

\author{M. Ostrowski}
\affil{Astronomical Observatory, Jagiellonian University, \\ul. Orla 171, 
30-244 Krak\'ow, Poland}

\begin{abstract} We report new detections of the hotspots in Cygnus A at
4.5 and 8.0 microns with the {\it Spitzer Space Telescope}.  Together with
detailed published radio observations and synchrotron self-Compton
modeling of previous X-ray detections, we reconstruct the underlying
electron energy spectra of the two brightest hotspots (A and D).  The
low-energy portion of the electron distributions have flat power-law
slopes ($s \sim1.5$) up to the break energy which corresponds almost
exactly to the mass ratio between protons and electrons; we argue that
these features are most likely intrinsic rather than due to absorption
effects. Beyond the break, the electron spectra continue to higher
energies with very steep slopes $s >$3. Thus, there is no evidence for the
`canonical' s=2 slope expected in 1st order Fermi-type shocks within the
whole observable electron energy range. We discuss the significance of
these observations and the insight offered into high-energy particle
acceleration processes in mildly relativistic shocks.  \end{abstract}

\section{Introduction}

Cygnus A is the nearest example ($z$=0.056) of a powerful Fanaroff-Riley
type-II radio galaxy. As such, it contains some of the brightest radio
hotspots to study these classical `working surfaces' of relativistic jets
(Hargrave \& Ryle 1974; Blandford \& Rees 1974). Although the hotspots are
well-studied at radio and X-ray wavelengths (\S~2), Cygnus A is viewed
through the galactic plane, making them difficult to study at optical
wavelengths (1.26 mag of extinction at V-band; Meisenheimer et al. 1997). 
Additionally, it is well known that bright foreground stars lie in the
direction of Cygnus A, and this has also hindered optical studies of the
hotspots. To circumvent these obstacles, we obtained new {\it Spitzer
Space Telescope} images (Figure~1) resulting in successful mid-infrared
detections of the hotspots. These new data, together with a wealth of
published high-quality measurements at other frequencies, give us some of
the most complete hotspot SEDs for a powerful radio galaxy. This
contribution is a summary of our recently published study (Stawarz et al.
2007).

\begin{figure}[ht] \plotone{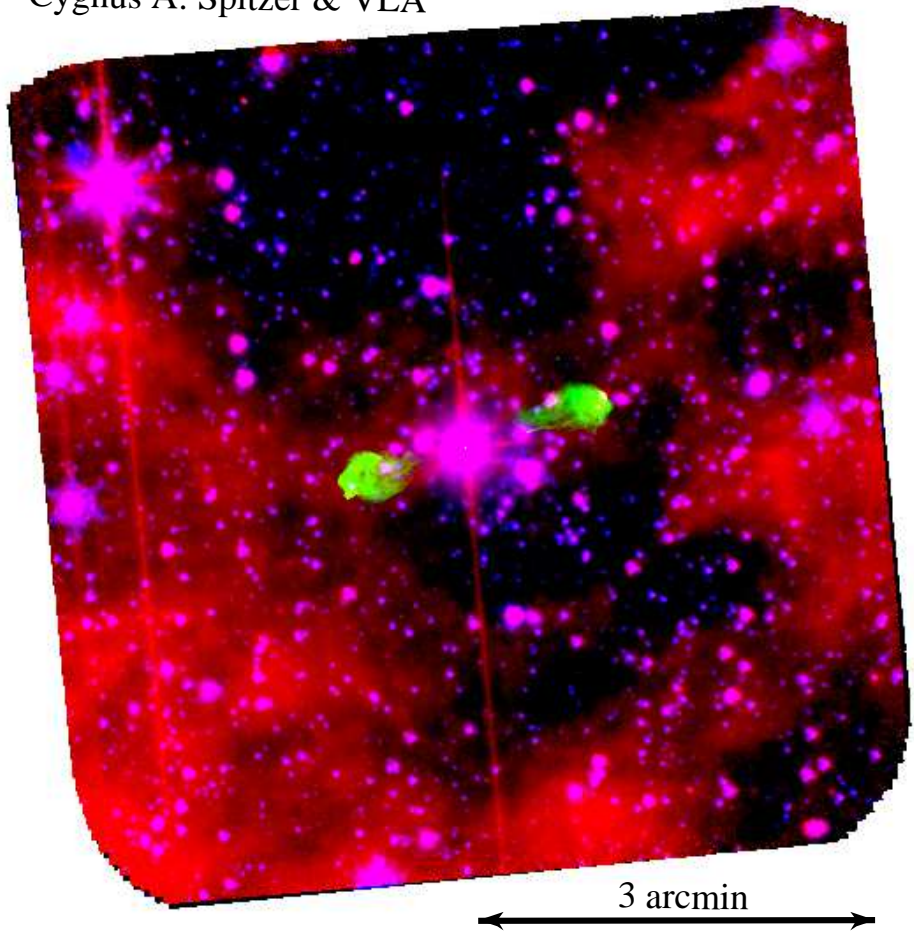} \caption{$Spitzer$
(our work) and VLA (green; from Perley et al. 1984) image of Cygnus A. The
4.5$\mu$m sources are shown in blue, 8.0$\mu$m emission in red, and those
emitting in both IR bands in pink.\label{fig-1}}\end{figure}

\section{$Spitzer$ Imaging and the Hotspot SEDs} 

\begin{figure}[ht] \plotone{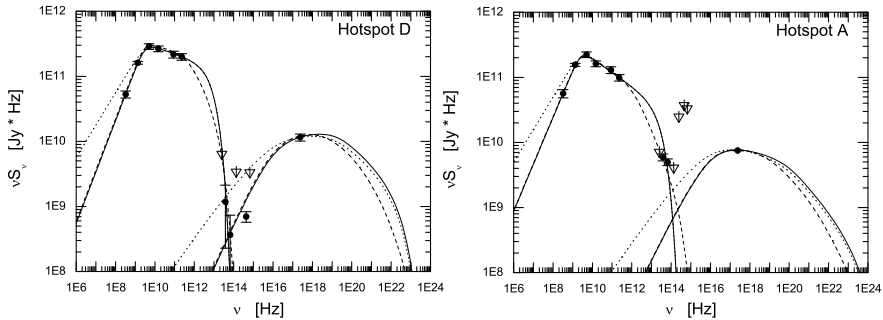} \caption{SEDs of the
two brightest radio hotspots (A, in the western lobe and D in the eastern
one) in Cygnus A.  The data points are shown with error bars and open
triangles are upper limits. The solid lines are our fits to the
synchrotron and SSC components of the SEDs. Dashed lines show the effect
of adding an additional break in the synchrotron spectrum ($\nu_{\rm br}$
= 0.5$\times$10$^{12}$ Hz for D and 1.2$\times$10$^{12}$ Hz for A). The
dotted lines correspond to a hypothetical increase of $\alpha_{\rm 1}$ to
0.5. \label{fig-2}}\end{figure}

We obtained new $Spitzer$/IRAC \citep{faz04} observations simultaneously
at 4.5 and 8.0 $\mu$m for 3.6 hrs. Galactic extinction amounts to only
0.02--0.03 mag at these bands. A foreground star is known to lie within an
arcsecond of the position of hotspot A -- extrapolating its spectrum into
the IRAC bands, its contamination to the hotspot IR fluxes is expected to
be negligible (see Fig.~2g in Meisenheimer et al. 1997). Radio data were
obtained from published VLA and BIMA measurements and maps
\citep{car91,wri04,laz06}. We focus our discussion on the two brightest
hotspots (A and D), whose spectral energy distributions (SEDs) are shown
in Figure~\ref{fig-2}.

We modeled the synchrotron (radio-to-infrared) portion of the hotspot SEDs
with a double power-law that decays exponentially at the higher
frequencies: 

\begin{equation}
S_{\nu}^{\rm syn} \propto \left\{ \begin{array}{ccc} \nu^{-\alpha_1} & {\rm 
for} & \nu_{\rm min} < \nu < \nu_{\rm cr} \\
\nu^{-\alpha_2} \, \exp\left(-\nu/\nu_{\rm max}\right) & {\rm for} & \nu > 
\nu_{\rm cr}
\end{array} \right. \quad .
\end{equation}

\smallskip \noindent The fitted parameters ($\alpha_1$, $\alpha_2$,
$\nu_{\rm cr}$, $\nu_{\rm max}$) are (0.28, 1.2, 2.6$\times$10$^{9}$ Hz,
$>$3.3$\times$10$^{13}$ Hz) for hotspot A, and (0.21, 1.1,
3.1$\times$10$^{9}$ Hz, 0.9$\times$10$^{13}$ Hz) for hotspot D. The
minimum frequency, $\nu_{\rm min}$, can be extrapolated down to 74 MHz
with the VLA detections of the hotspots by Lazio et al. (2006) at this
frequency. 

The hotspots emit strong X-ray emission by the synchrotron self-Compton
(SSC) process as discovered with $ROSAT$ (Harris et al. 1994). Assuming
this for the $Chandra$ detection of the hotspots (Wilson et al. 2000; 
Wright \& Birkinshaw 2004), we infer magnetic field strengths of
$\sim$0.2--0.3 mG. The emitting plasma is close to equipartition with
$U_{e}/U_{B}=3-8$ (Figure~3), consistent with the previous fits of the
$Chandra$ data.  An $HST$ optical measurement of hotspot D (Nilsson et al.
1997) lies near the low energy extrapolation of the SSC spectrum; this
makes it one of the few hotspots with detected optical SSC emission
(Brunetti 2004).  Increasing the low energy spectral index ($\alpha_1$) of
hotspot D from its observed value to the `canonical' value of 0.5 would
overproduce the observed optical and IR emissions, so the observed flat
spectra are likely intrinsic.

\begin{figure}[ht] \plotone{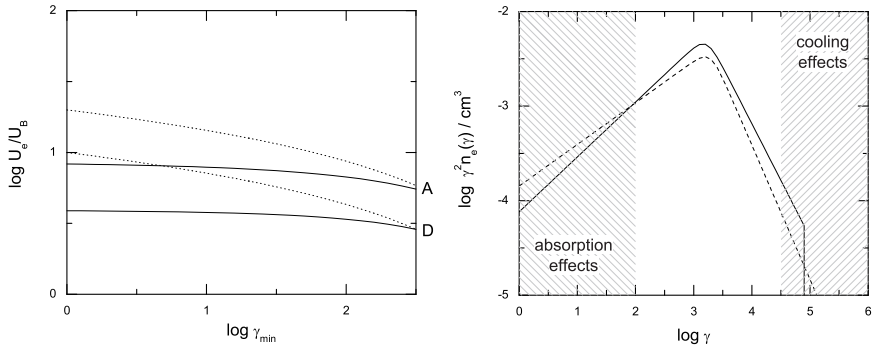} \caption{[left]
Ratio of energy densities in ultrarelativistic electrons and magnetic
field for the two brightest radio hotspots as a function of the low energy
cutoff in $n_{\rm e}(\gamma)$. Solid lines correspond to the case with
observed low-frequency synchrotron spectral indices of $\alpha_1 = 0.21$
(hotspot D; lower line) and $0.28$ (hotspot A; upper line), while dotted
lines correspond to a hypothetical increase of $\alpha_1$ to 0.5 for both
features. [right] Electron energy spectrum for hotspots A (dashed line)
and D (solid line), as inferred from the SSC modeling. Shaded regions show
the energy ranges affected by absorption or radiative losses.
\label{fig-3}}\end{figure}

With the magnetic field and observed spectral fits, we can reconstruct the
underlying electron energy spectra, $n_{\rm e}(\gamma) \propto
\gamma^{-s}$ ($s=2\alpha+1$; Figure~3). They have the following features: 
{\it (i) The critical break energy corresponds almost exactly to the ratio of
the proton to electron masses, $\gamma_{\rm cr}\simeq m_{\rm p}/m_{\rm
e}$. (ii) The low-energy segment of the electron distributions continues down
to at least $\sim$0.1 $m_{\rm p}/m_{\rm e}$, with flat spectra; the
power-law slopes are $s\simeq$1.4--1.6. (iii) The new IR data constrain the high-energy electron spectra, with
indices $s$$>$3 and maximum energies $\simgt$50 $m_{\rm p}/m_{\rm e}$.}
The ``standard'' electron spectrum expected from the diffusive
(1st order Fermi) shock acceleration in the non-relativistic test-particle
limit, $n_{\rm e}(\gamma) \propto \gamma^{-2}$, is not observed. The
overall curved spectra are likely to be intrinsic rather than due to any
absorption/cooling effects.

\section{The Cygnus A Hotspots as Relativistic Shocks}

Terminal shocks of powerful jets, like those in Cygnus A, are probably
(mildly) relativistic, with an oblique magnetic field configuration.  As
such, it should not be surprising that the diffusive acceleration process
known from the non-relativistic test-particle models are not directly
applicable, i.e., the lack of a canonical shock-type spectrum $\propto
\gamma^{-2}$ -- see Begelman \& Kirk (1990) and
Niemiec et al. (2006) for relevant discussions.  The break frequency
corresponding to the $m_{\rm p}/m_{\rm e}$ mass ratio preceded by a
flat-spectrum power-law is also an interesting finding, which confirms
some previous evidences for such in other powerful hotspots (see Leahy et al. 1989). 

Our results, although inconsistent with the simplest modeling of terminal
hotspots in radio galaxies as sites of the efficient 1st order Fermi
acceleration, are instead in good agreement with our current understanding
of relativistic shock waves. In particular, a flat power-law electron
spectrum up to $m_{\rm p}/m_{\rm e}$ energies is consistent with the simple
(1D) resonant acceleration model discussed by Amato \& Arons (2006) for
relativistic shocks.  Their simulations show that as a result of resonant
absorption of cyclotron emission produced by cold protons reflected from
the shock front, a flat power-law electron energy distribution emerges
between energies $\gamma \sim \Gamma$ and $\Gamma \, (m_{\rm p}/m_{\rm
e})$, the latter possibly reduced by a factor of a few due to thermal
dispersion in the upstream proton momenta ($\Gamma$ is the bulk Lorentz
factor of the upstream medium). The slope of this distribution is not well
constrained by the simulations, but can be relatively flat and appears to
depend on the plasma content (i.e., on the ratio of proton to electron
number density). Although these particular simulations involve a large
$\Gamma$ plasma, analogous processes are expected to take place also at
mildly relativistic shocks considered in hotspots like Cygnus A ($\Gamma
\sim$ few), so the model results can at least be applied qualitatively. 

If such an association is correct, and if the results of Amato \& Arons
(2006) are applicable, then it would automatically imply that ({\it i})
the jets in powerful radio sources like Cygnus~A are made of an
electron-proton rather than electron-positron plasma, ({\it ii}) a
significant fraction of the jet kinetic power is carried by cold protons, and
({\it iii}) the number density of protons within the jets is most likely
lower than that of the electron-positron pairs (see the discussion in
Stawarz et al. 2007). We note that these conclusions regarding the jet
content would then be consistent with those presented for quasar jets
(e.g., Sikora \& Madejski 2000), which are widely regarded as beamed
counterparts of FR~II radio galaxies like Cygnus A.

\acknowledgements 

This work is based on observations made with the {\it Spitzer Space
Telescope}, which is operated by the Jet Propulsion Laboratory, California
Institute of Technology under a contract with NASA.  The NRAO is operated
by Associated Universities, Inc. under a cooperative agreement with the
National Science Foundation. \L.~S. and M.~O. were supported by MEiN
through the research project 1-P03D-003-29 in years 2005-2008. The work at
SAO was partially supported by {\it Spitzer} grant 1279229.

\end{document}